\documentclass[aps,preprint,groupedaddress]{revtex4-1}
\usepackage{mathtools}
\usepackage{amsmath}
\usepackage{amsfonts}
\usepackage{amssymb}
\usepackage{graphicx}
\usepackage{epstopdf}

\newcommand*{\eqb}{\begin{equation}}
\newcommand*{\eqe}{\end{equation}}

\newcommand{\abs}[1]{\left| #1 \right|}

\newcommand{\R}{\mathbb{R}}

\newcommand{\pd}[1]{\frac{\partial}{\partial{#1}}}

\begin{document}

% Use the \preprint command to place your local institutional report
% number in the upper righthand corner of the title page in preprint mode.
% Multiple \preprint commands are allowed.
% Use the 'preprintnumbers' class option to override journal defaults
% to display numbers if necessary
%\preprint{}

%Title of paper
\title{Comment on ``Fokker-Planck equations for nonlinear dynamical systems driven by non-Gaussian L\'evy processes'' [J. Math. Phys.  53, 072701 (2012)]}

% repeat the \author .. \affiliation  etc. as needed
% \email, \thanks, \homepage, \altaffiliation all apply to the current
% author. Explanatory text should go in the []'s, actual e-mail
% address or url should go in the {}'s for \email and \homepage.
% Please use the appropriate macro foreach each type of information

% \affiliation command applies to all authors since the last
% \affiliation command. The \affiliation command should follow the
% other information
% \affiliation can be followed by \email, \homepage, \thanks as well.
\author{Marcin Magdziarz}

\email[]{marcin.magdziarz@pwr.edu.pl}
%\homepage[]{Your web page}
%\thanks{}
%\altaffiliation{}
\affiliation{Hugo Steinhaus Center, Faculty of Pure and Applied Mathematics, Wroclaw University of Technology,  ul. Wyspianskiego 27,
50-370 Wroclaw, Poland}
\author{Tomasz Zorawik}
\email[]{tomasz.zorawik@pwr.edu.pl}
%\homepage[]{Your web page}
%\thanks{}
%\altaffiliation{}
\affiliation{Hugo Steinhaus Center, Faculty of Pure and Applied Mathematics, Wroclaw University of Technology,  ul. Wyspianskiego 27,
50-370 Wroclaw, Poland}

%Collaboration name if desired (requires use of superscriptaddress
%option in \documentclass). \noaffiliation is required (may also be
%used with the \author command).
%\collaboration can be followed by \email, \homepage, \thanks as well.
%\collaboration{}
%\noaffiliation

\date{\today}

\begin{abstract}
In an article [J. Math. Phys.  53, 072701 (2012)] X. Sun and J. Duan presented Fokker-Planck equations for nonlinear stochastic differential equations with non-Gaussian L\'evy processes. In this comment we show a serious drawback in the derivation of their main result. In the proof of Theorem 1 in the aforementioned paper, a false assumption that each infinitely differentiable function with compact support is equal to its Taylor series, is used. We prove that although the derivation is incorrect, the result remains valid only if we add certain additional assumptions.
\end{abstract}

% insert suggested PACS numbers in braces on next line
%\pacs{}
% insert suggested keywords - APS authors don't need to do this
%\keywords{}

%\maketitle must follow title, authors, abstract, \pacs, and \keywords
\maketitle

% body of paper here - Use proper section commands
% References should be done using the \cite, \ref, and \label commands
%\section{Introduction}
X. Sun and J. Duan analyzed in \cite{article} the following It\^o stochastic differential equation:
\eqb
\label{ito}
dX_t=f(X_t,t)dt+\sigma(X_{t-},t)dL_t,\quad X_s=x,
\eqe
where $L_t$ is a L\'evy process with L\'evy triplet $(b,A,\nu)$, $f$ - drift and $\sigma$ - noise intensity. Under the assumption that $f$ and $\sigma$ satisfy Lipschitz and growth conditions the authors proved the following theorem:

\textbf{Theorem 1 \cite{article}}: \textit{
The Fokker-Planck equation for the It\^o SDE (\ref{ito}) is
\eqb
\begin{split}
&\frac{\partial p(y,t;x,s)}{\partial t}=-\pd{y}(\rho(y,t)p(y,t;x,s))+\frac{1}{2}A\frac{\partial^2}{\partial y^2}(\sigma^2(y,t)p(y,t;x,s))\\
&\quad\quad+\int_{\R\setminus\{0\}}\bigg[\sum_{k=1}^\infty\frac{(-z)^k}{k!}\frac{\partial^k}{\partial y^k}(\sigma^k(y,t)p(y,t;x,s))+I_{(-1,1)}(z)z\frac{\partial}{\partial y}(\sigma(y,t)p(y,t;x,s))\bigg] \nu (dz),
\end{split}
\eqe
where $\rho(x,t)=f(x,t)+b\sigma(x,t)$ and $I_{(-1,1)}(x)$ is the indicator function of the set $(-1,1)$.
}
However, there is a serious error in the derivation of this result, which makes the whole proof wrong. X. Sun and J. Duan used the Taylor expansion to obtain in Eq. (30) in \cite{article} that
\eqb
\phi(x+y\sigma(x,t))=\phi(x)+\sum_{k=1}^\infty\frac{y^k}{k!}\sigma^k(x,t)\frac{\partial^k}{\partial x^k}\phi(x)
\eqe
for $\phi(x)\in C_0^\infty(\R)$ - the space of smooth functions with compact supports. The problem with this equation is that the only $f\in C_0^\infty(\R)$ which is equal to its Taylor series  is the constant function $f \equiv 0$ - see Corollary 1.2.5 in \cite{real analytic}.
This error has serious consequences in the further reasoning in \cite{article}. In the proof it was necessary to find an adjoint operator of the following operator (Eq. (27) in \cite{article})
\eqb
\label{operator}
A_{2t}\phi(x)=\int_{\R\setminus\{0\}}\bigg[\phi(x+y\sigma(x,t))-\phi(x)-I_{(-1,1)}(y)y\sigma(x,t)\pd{x}\phi(x)\bigg]\nu(dy),
\eqe
where $\phi(x)\in C_0^\infty(\R)$. X. Sun and J. Duan in Eq. (32) in \cite{article} claimed that the operator $A_{2t}$ can be represented as
\eqb
\label{operator transformed}
\begin{split}
A_{2t}\phi(x)=\int_{\R\setminus\{0\}}\bigg[\sum_{k=1}^\infty\frac{y^k}{k!}\sigma^k(x,t)\frac{\partial^k}{\partial x^k}\phi(x)-I_{(-1,1)}(y)y\sigma(x,t)\frac{\partial}{\partial x}\phi(x) \bigg]\nu (dy)
\end{split}
\eqe
However, this is not always true. Below we present a counterexample. Let us take $\nu=\delta_1$ - Dirac delta concentrated at $z=1$. This L\'evy measure corresponds to Poisson process $L(t)$ with the rate $\lambda=1$ (see \cite{applebaum}). We also take $\sigma(x,t)=-x$ which obviously satisfies Lipschitz and growth conditions.
Then Eq. (\ref{operator}) gives us
\eqb
A_{2t}\phi(x)=\phi(0)-\phi(x),
\eqe
whereas Eq. (\ref{operator transformed}) now has the form
\eqb
A_{2t}\phi(x)=\sum_{k=1}^\infty\frac{(-x)^k}{k!}\frac{\partial^k}{\partial x^k}\phi(x).
\eqe
These two forms are not equivalent. For instance, if we set $\phi(x)=\exp\left(-\frac{1}{1-x^2}\right)1_{(-1,1)}(x)$ which satisfies $\phi\in C_0^\infty(\R)$ then $A_{2t}\phi(x)$ given by the second form vanishes outside the interval $(-1,1)$, whereas the first one does not.

One can try to correct this error by changing the space of the test functions. Instead of $C_0^\infty(\R)$ one can take analytic functions which, together with all their derivatives, decay fast enough at infinity. For instance one can take the functions of the form $\phi(x)=\exp\left(-\frac{(x-a)^2}{b}\right)$, where $a\in \R$ and $b\in\R^+$. For these functions one can use their Taylor expansions. However, it is still necessary to justify one of the transformations in Eq. (34) in \cite{article}:
\eqb
\begin{split}
&\int_\mathbb{R}\bigg[\sum_{k=1}^\infty\frac{z^k}{k!}\sigma^k(y,t)\frac{\partial^k}{\partial y^k}\phi(y)-I_{(-1,1)}(z)z\sigma(y,t)\frac{\partial}{\partial y}\phi(y) \bigg]p(y,t;x,s) dy\\
&=\int_\mathbb{R}\bigg[\sum_{k=1}^\infty\frac{(-z)^k}{k!}\frac{\partial^k}{\partial y^k}(\sigma^k(y,t)p(y,t;x,s))+I_{(-1,1)}(z)z\frac{\partial}{\partial y}(\sigma(y,t)p(y,t;x,s))\bigg] \phi(y)dy
\end{split}
\eqe
for all $z\in \R\setminus\{0\}$. We have
\eqb
\begin{split}
\label{interchange}
&\int_\mathbb{R}\bigg[\sum_{k=1}^\infty\frac{z^k}{k!}\sigma^k(y,t)\frac{\partial^k}{\partial y^k}\phi(y)-I_{(-1,1)}(z)z\sigma(y,t)\frac{\partial}{\partial y}\phi(y) \bigg]p(y,t;x,s) dy\\
&=\sum_{k=1}^\infty\int_\mathbb{R}\bigg[\frac{z^k}{k!}\sigma^k(y,t)\frac{\partial^k}{\partial y^k}\phi(y)-I_{(-1,1)}(z)z\sigma(y,t)\frac{\partial}{\partial y}\phi(y) \bigg]p(y,t;x,s) dy\\
&=\sum_{k=1}^\infty\int_\mathbb{R}\bigg[\frac{(-z)^k}{k!}\frac{\partial^k}{\partial y^k}(\sigma^k(y,t)p(y,t;x,s))+I_{(-1,1)}(z)z\frac{\partial}{\partial y}(\sigma(y,t)p(y,t;x,s))\bigg] \phi(y)dy.
\end{split}
\eqe
Now, we want to interchange the integral with the sum, but this is not always possible. The series
\eqb
\label{series}
\sum_{k=1}^\infty\frac{(-z)^k}{k!}\frac{\partial^k}{\partial y^k}(\sigma^k(y,t)p(y,t;x,s))
\eqe
can diverge. One way of dealing with this problem is to add an  assumption to Theorem 1 that the interchange of the integral and the sum in Eq. (\ref{interchange}) is allowed. Another possible solution is to
assume that $\sigma(x,t)=\sigma(t)$ and $p(y,t;x,s)$ satisfies $\abs{\frac{\partial^k}{\partial y^k}p(y,t;x,s)}<MC^k$ where $M>0$ and $C>0$. In such situation  the interchange of operators is possible based on the Dominated convergence theorem, see for example \cite{magdziarz zorawik}.
% If you have acknowledgments, this puts in the proper section head.
\begin{acknowledgments}
This research was partially supported by the Ministry of Science and Higher Education of Poland
program Iuventus Plus no. IP2014 027073.
\end{acknowledgments}

% Create the reference section using BibTeX:
%\bibliography{basename of .bib file}

\end{document}